\documentclass[conference]{IEEEtran}
\usepackage{amsmath}
\usepackage{amssymb}
\usepackage{floatfig,epsfig}
\usepackage{subfigure}
\usepackage{verbatim}
\usepackage{pifont}
\usepackage{algorithm}
\usepackage{algorithmic}
\usepackage{alltt}
\usepackage{bm}
\usepackage{isolatin1}
\usepackage{graphics}
\usepackage{epic}
\usepackage{eepic}
\usepackage{graphpap}
\usepackage{cite}
\usepackage{psfrag}

\newtheorem{pro}{Proposition}

\ifCLASSINFOpdf
\else
\fi

\begin{document}

\title{Repair for Distributed Storage Systems with Erasure Channels}

\author{Majid Gerami, Ming Xiao
\\  Communication Theory Lab, Royal Institute of Technology, KTH, Sweden, \\ E-mail: \{gerami, mingx\}@kth.se\\}
\date{\today}


\maketitle

\begin{abstract}
\normalsize
We study the repair problem of distributed storage systems in erasure networks where the packets transmitted from surviving nodes to the new node might be lost. The fundamental storage-bandwidth tradeoff is calculated by multicasting analysis in erasure networks.
 The optimal tradeoff bound can be asymptotically achieved when the number of transmission (packets) goes to infinity. For a limited number of transmission,
 we study the probability of successful regenerating. Then, we investigate two approaches of increasing the probability of successful regenerating, namely, by connecting more surviving nodes or by increasing the storage space of nodes.  Using more nodes may pose larger delay and in certain situation
 it might not be possible to connect to more nodes too. We show that in addition to reducing repair bandwidth, increasing storage space can
also increase reliability for repair.


\end{abstract}

%

\IEEEpeerreviewmaketitle

\section{Introduction}\label{sec:intro}
\begin{figure}
\centering
\psfrag{s}[][][3.5]{ $S$ }
\psfrag{A}[][][3.5]{DC }
\psfrag{node1}[][][2.5]{node 1 }
\psfrag{node2}[][][2.5]{node 2 }
\psfrag{node3}[][][2.5]{node 3 }
\psfrag{node4}[][][2.5]{node 4 }
  \psfrag{mk}[][][3.0]{$M/k$ }
\resizebox{8cm}{!}{\epsfbox{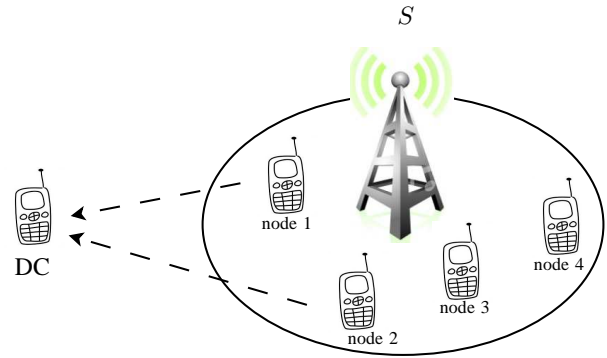}}
\caption{A distributed storage system in a wireless  network. Base station distributes a file among storage nodes on its coverage area. A data collector (DC) can recover the source file, even though it is out of base station coverage, after meeting certain number of storages node. In this example, the source file is coded by ($n=4, k=2$)-MDS code. Thus, the DC by downloading from  any two  storage nodes can rebuild the source file.}
\label{Fig.1WirelessDTN}
\end{figure}

Distributed storage systems have attracted substantial research interests recently for increasing applications in e.g., file sharing, cloud storage. The main benefit of distributing a file within a network is to provide reliable storage using unreliable storage nodes. Distributed storage systems also bring applications in wireless networks where links between storage nodes may be  unreliable. For instance, consider an application of  distributed storage systems in a \textit{delay tolerant network} \textit{(DTN)} with wireless channels, as shown in  Fig. \ref{Fig.1WirelessDTN}. A DTN is a kind of networks in which there might not be direct links between a source and destinations and  services can tolerate an acceptable level of incurred delay \cite{Jain01}. In Fig. \ref{Fig.1WirelessDTN}, a base station distributes a source file at the mobile nodes which may travel towards arbitrary directions.  Then a data collector (DC) by meeting  a certain number of these mobile storage nodes can rebuild the source file.


Unreliability in distributed storage systems might stem from disk failure, but it is not limited to this reason. For example, in the network in Fig.  \ref{Fig.1WirelessDTN}, a node failure may be because a mobile node leave the system. In distributed storage systems in wireless networks,  unreliability may also occur in a transmission process due to link failure, congestion, or buffer overflow and so on. We model an unreliable link in the network to an erasure channel. When a packet erasure happens, all the information in the packet is lost.

In a distributed storage system, a source file is generally encoded by erasure codes and distributed on redundant nodes. Thus, the file is still available even though some storage nodes fail. The maximum reliability by using erasure codes is achieved when a file contains $k$ fragments and is encoded to $n$ fragments, and every $k$ out of $n$ fragments can rebuild the original file. The codes are known as MDS (Maximum Distance Separable) codes. Then, an $(n,k)$ erasure code has the MDS property if every $k$ out of $n$ fragments can rebuild the original file. For instance, if a file of size $M$ is divided into $k$ fragments and encoded by $(n,k)$ MDS ($(4, 2)$ here) codes on the application of Fig. \ref{Fig.1WirelessDTN}, then a receiver (or data collector) by observing at least $k, (k=2)$ storage nodes can rebuild the source file.

To maintain the reliability of the system, when a node fails, a new node should be regenerated.  In the regenerating process, surviving nodes transmit sufficient data to the new node such that the new node can maintain the MDS property, but the new node may have different coded symbols.  This is called the functional repair.  We only consider the functional repair in the paper.
A naive approach for repair  is to rebuild the source file (by download $M$ fragments from surviving nodes) and then regenerate the new node with the MDS property. However, downloading $M$ fragments for regenerating $M/k$ fragments is not efficient in term of bandwidth. Reference \cite{Dimk01} introduces network coding to distributed stroage systems in error-free networks (\textit{lossless network}) and models the repair process by an information flow graph into a multicast problem. Moreover, it is shown that by slightly increasing node storage,  the bandwidth can be reduced considerably \cite{Dimk01}. In \cite{Dimk01}, the fundamental bandwidth-storage tradeoff was established. The codes on the optimum tradeoff are called  \textit{regenerating codes}.

Two extreme scenarios on the fundamental bandwidth-storage tradeoff are \textit{minimum storage regenerating (MSR)} and \textit{minimum bandwidth regenerating (MBR)} codes. For the MSR code, like the MDS codes, a file of size $M$ is divided into $k$ fragments and then encoded to $n$ fragments. Hence, each node stores $M/k$ fragments.  Unlike MDS codes, MSR codes have the minimum bandwidth in regenerating a new node. That is, for $d$ number of surviving nodes joining in the repair process, the minimum bandwidth by each surviving node ($\beta$) is calculated by
\begin{eqnarray}
\alpha_{MSR}=&M/k, \nonumber  \\
\beta_{MSR}=&\frac{M}{k(d-k+1)}.
\label{Eq-MSR}
\end{eqnarray}

For  MBR codes,  each node stores  more fragments than $M/k$ but used repair bandwidth is exactly the same as node storage, and is calculated by
\begin{eqnarray}
\alpha_{MBR}=&\frac{2dM}{k(2d-k+1)}, \nonumber  \\
\beta_{MBR}=&\frac{2M}{k(2d-k+1)}.
\label{Eq-MBR}
\end{eqnarray}

After seminal work in \cite{Dimk01}, references \cite{Wu01} studies the code construction and achievablity of the functional and exact repair. In \cite{Ken01}, cooperative regenerating codes are considered to reduce the bandwidth in the scenario of multiple-node failure. Reference \cite{Maj01} suggests surviving nodes cooperation in order to minimizes the cost of repair in  multi-hop networks. Yet, in most of previous work of regenerating codes, it is assumed that links between storage nodes are perfect, without any error and erasure. In distributed storage systems especially in the wireless networks, as the example in Fig. 1, packets on the channels might get lost. Then, the redundant data needs to be transmitted for repair. Recently \cite{Rashmi02} has suggested a regenerating code which is resistant to a specific number of path failures by requesting more nodes to join the repair process. Particularly, in their approach the code resistant to $t$ number of path failure, it is required to transmit from $d^{'}=d+t$ surviving nodes instead of $d$ nodes (that for perfect channels). However, reference \cite{Rashmi02} has not studied the probability of successful regenerating and how to construct the optimal (in regard of probability of successful regenerating) codes. We call a regenerating process is successful when the new node beside surviving nodes has MDS property. We shall consider the probability of successful regenerating in the repair process and show that finding the optimum values for $d, \text{ and } d^{'}$ depends on the erasure probability of the links. Thus, we can find the code maximizing the probability of successful regenerating, given the constraints of bandwidth and storage capacity. We also find that by using slightly more storage, the probability of successful regenerating can be further increased. This approach suits for applications when requesting more surviving storage nodes may cause large delay in the repair or when there is not possible to connect to more storage nodes.

Other related works on network coding for erasure networks are as follows.  The capacity of wireless erasure  networks has been studied in \cite{Dana01}. In references \cite{Jain02},\cite{Leong01}, the probability of successful reconstruction of a source file is studied in erasure networks. However, the papers do not studied the probability of successful regenerating. References \cite{Jain01}, \cite{Jain02} study approaches in delay tolerant networks. In \cite{Jain02}, the author show that there is no unique answer to the question whether coding  for distributing a file  maximizes the probability of successful reconstruction or not.

The rest of this paper is orangized as follows. In Section \ref{sec:Tradoff} we study the fundamental bandwidth-storage tradeoff in erasure networks. In Section \ref{sec:OptimumCode} we study the maximum probability of successful regenerating. In Section \ref{sec:addstorage} we propose to use more storage for increasing the probability of successful repair. Finally we conclude the paper in Section \ref{sec:conclusion}.

\section{Fundamental bandwidth-storage tradeoff in distributed storage networks with erasure channels}\label{sec:Tradoff}

In \cite{Dimk01}  the fundamental bandwidth-storage tradeoff was found by cut analysis in a  multicast network where links have no error or erasure (lossless networks). For the  multicast problem in erasure networks, the capacity of networks was found in \cite{Dana01}. Reference \cite{Dana01} proved the multicast capacity can be achieved in erasure networks by linear network coding. Using similar approaches to \cite{Dimk01}, \cite{Dana01}, we can find the fundamental tradeoff of distributed storage systems in erasure networks.
 To have a closed-form formula for the fundamental bandwidth-storage tradeoff with erasure channels, we consider the scenario when all the links in the network have equal erasure probabilities $\varepsilon$. Then, we formulate the tradeoff as follows.

\begin{figure}
 \centering
 \psfrag{data1data1data1}[][][2.5]{ $\varepsilon=0.1$ }
 \psfrag{data2data2data2}[][][2.5]{ $\varepsilon=0.2$ }
 \psfrag{data3data3data3}[][][2.5]{ $\varepsilon=0.3$ }
 \psfrag{ylabel}[][][1.5]{ $\alpha$ }
 \psfrag{xlabel}[][][1.5]{ $\gamma$ }
 \resizebox{8cm}{!}{\epsfbox{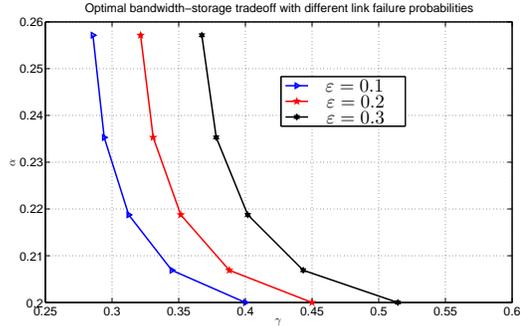}}
\caption{Fundamental bandwidth-storage tradeoff in erasure networks, $n=10,k=5$. Higher link failure probability ($\varepsilon$) poses higher traffic on the repair.}
 \label{tardeoffs}
\end{figure}

\begin{pro} \label{Proposition:Regbound}
Consider a repair process in a distributed storage system where channels from $d$ surviving nodes $(d \leq (n-1))$ to the new node have an erasure probability  $\varepsilon$. For any $\alpha \geq \alpha^{\ast}(M,n,k,d,\gamma,\varepsilon)$ there is a linear network code to achieve the point $(n,k,d,\alpha,\gamma,\varepsilon)$, and for $\alpha < \alpha^{\ast}(M,n,k,d,\gamma,\varepsilon)$, it is impossible to find a code. Here $n$ denotes the number of storage nodes, and every $k$ storage nodes can rebuild the source file. $d$ is the number of surviving nodes participating the repair, and $\alpha$ is each node storage, and  $\gamma$ is total repair bandwidth.  Therefore, the fundamental bandwidth-storage tradeoff can be calculated as,
\begin{eqnarray}
\alpha^{\ast}=
\begin{cases} M/k & \text{  if    } \gamma \in [\frac{f(0)}{1-\varepsilon},+ \infty) \\
\frac{M-g(i)(1-p)\gamma}{k-i} & \text{  if    } \gamma \in [\frac{f(i)}{1-\varepsilon},\frac{f(i-1)}{1-\varepsilon}), \end{cases}
\label{Eq-Regbound}
\end{eqnarray}
\end{pro}
where,
\begin{equation}
f(i)=\frac{2Md}{(2k-i-1)i+2k(d-k+1)},
\end{equation}
and,
\begin{equation}
g(i)=\frac{(2d+2k+i+1)i}{2d}.
\end{equation}
\begin{proof}The proof is similar to that in \cite{Dimk01}. The only difference is that in the erasure network the new node in average receives $(1-\varepsilon)\beta $ from each surviving node. \end{proof}

Fig. \ref{tardeoffs} shows the fundamental bandwidth-storage tradeoffs when  links have erasure probabilities $\varepsilon=0.1,0.2,0.3$, respectively. As we expect, larger erasure probabilities lead to higher traffic in the repair process.
These optimum bandwidth-storage tradeoff can be achieved asymptotically, i.e., when the number of transmission  packets goes to infinity. When feedback is available on the network the optimal bound can be achieved by retransmission (potentially infinite). However, for a network without feedback,   rateless random linear network codes \cite{Lun01} can asymptotically achieve the optimal tradeoff.
For a limited number of transmission (retransmission) we are interested to maximize the probability of successful repair.  We note that a repair process is called of success when the new node has the MDS property (along with surviving nodes).

\section{Maximizing the probability of successful regenerating $p_s$} \label{sec:OptimumCode}

Obviously, one approach to provide redundancy (for erasure) for repair is to connect more surviving nodes to the new node. That is, for the repair in a lossless network $\gamma=d\beta$ packets are transmitted from $d$ nodes (each node transmits $\beta$ fragments).  Then in an easure network, $d^{'}$ ($d^{'}\geq d$) nodes transmitting $\gamma^{'}=d^{'}\beta$ packets may be necessary. One interesting scenario is that with the constraint of the total repairing bandwidth in the network $(\gamma^{'} \leq \gamma_{th})$, how to find the optimal repair (maximizing the probability of successful regenerating $p_s$). We will show that solving the problem  depends on the link erasure probability.

To formulate the problem, consider a repair process in a distributed storage system. Links from surviving nodes to the new node fails with  probability $\varepsilon$. To simplify illustration, we assume all the links have equal erasure probabilities.
To combat link erasures, the redundant data should be transmitted. In a lossless network for  regenerating a new node with parameters $(n,k,d,\alpha,\gamma=d\beta)$. By receiving $\gamma=d\beta$ data, the new node is regenerated successfully.  Therefore, a regenerating process is successful when all $\gamma=d\beta$ data are received. However, for an erasure networks,  failure at links  can cause the repair unsuccessful. Thus, to increase the probability of successful regenerating, one approache is to use $d^{'}$ $(d^{'}\geq d)$ surviving nodes for repair.  Each surviving node still transmits $\beta$ fragments of data. Hence,  the probability of successful regenerating is that the new node receives  from at least $d$  out of $d^{'}$ surviving nodes, which is calculated by
 \begin{equation}
 p_s=\sum_{i=d}^{d^{'}} \binom{d^{'}}{i} (1-\varepsilon)^i \varepsilon^{d^{'}-i}.
 \end{equation}

\begin{figure}
 \centering
 \psfrag{data1data1data1data1}[][][1.5]{ $d=7,d^{'}=9,\gamma^{'}=3.6$ }
 \psfrag{data2data2data2data2}[][][1.5]{ $d=5,d^{'}=6,\gamma^{'}=4$ }
 \psfrag{title}[][][1.5]{Probability of successful regenerating in a DSS with n=10, k=5}
  \psfrag{ylabel}[][][1.5]{ $p_s$ (successful regenerating probability)  }
 \psfrag{xlabel}[][][1.5]{ $\varepsilon$ (erasure probability) }
\resizebox{8cm}{!}{\epsfbox{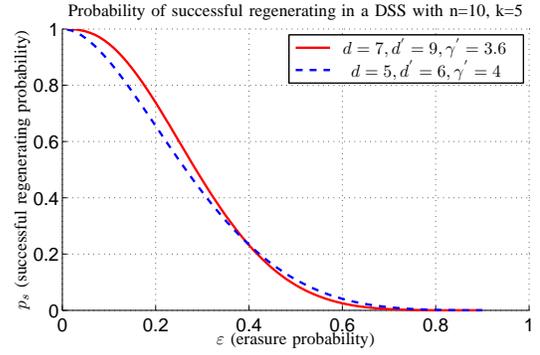}}
\caption{A scheme with a higher probability of success in one region might has a smaller probability of success in other region.}
 \label{Fig.ps_pc}
\end{figure}

In Fig. \ref{Fig.ps_pc}, the probability of successful regenerating with various erasure probabilities using two  regenerating schemes has been shown. In the example, the distributed storage system contains $n=10$ storage nodes and every $k=5$ nodes can rebuild the source file.  In the first scheme a minimum-bandwidth regenerating (MBR) code with parameters $d=7, d^{'}=9$ is used. In the second scheme a minimum-bandwidth regenerating  code with parameters
 $d=5,d^{'}=6$ is used. Using (\ref{Eq-MBR}), the transmitted bandwidth for the former is $\gamma^{'}=3.6$ and for the latter is  $\gamma^{'}=4$. From Fig.~\ref{Fig.ps_pc},  we can see that even though the bandwidth of the second scheme is higher than the first one, it has lower probabilities of success for $\varepsilon < 0.4$. This example motivates us to further investigate the problem of maximizing the probability of successful regenerating for the given repair bandwidth.

Given the constraint on the transmitted bandwidth $(\gamma^{'}=d^{'}\beta)$, $d$ and $d^{'}$ can be optimized to maximize the probability of successful repair. The optimization problem can be formulated as follows,

\begin{figure}
 \centering
 \psfrag{data1}[][][2.5]{ $d^{'}$ }
 \psfrag{data2}[][][2.5]{ $d$ }
 \psfrag{title}[][][1.5]{Maximizing $p_s$ for different integer values of $d$ and $d^{'}$}
  \psfrag{ylabel}[][][1.5]{ number of storage nodes  }
 \psfrag{xlabel}[][][1.5]{ $\varepsilon$ (erasure probability) }
\resizebox{8cm}{!}{\epsfbox{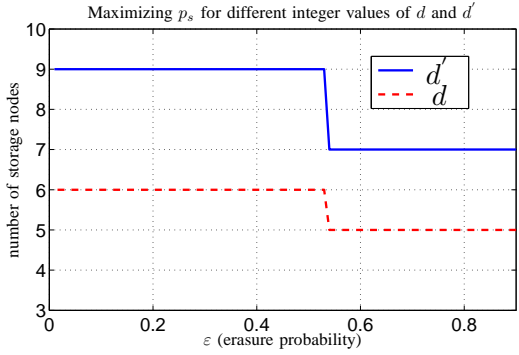}}
\caption{For a network with low $\varepsilon$ (erasure probability), maximizing $p_s$ uses a
larger number of surviving nodes for repair.  For a network with high $\varepsilon$, maximizing $p_s$  finds a smaller number of surviving nodes for repair.}
 \label{Fig.optimum_d_and_d2}
\end{figure}
 \begin{equation}
 \begin{array}{lc}
 \max\limits_{d,d'} &  p_s \\

 \mbox{\text{s. t.}} & d^{'}\beta \leq \gamma_{th} \\
 \end{array}
 \label{max-ps}
 \end{equation}

For illustration, we use the optimization problem in the previous example and find optimal $d$ and $d^{'}$ for given $\gamma_{th} \leq 5 $ over different erasure probabilities. The optimal  $d$ and $d^{'}$ with different erasure probabilities have been shown in Fig. \ref{Fig.optimum_d_and_d2}. We can see that for the network with high erasure probabilities, using fewer surviving nodes in the repair process maximizes $p_s$. Conversely, for the network with low erasure probability, using more surviving nodes  maximizes $p_s$.


\section{Increasing storage space of nodes}\label{sec:addstorage}

To provide redundancy, one approach is to use more surviving nodes for repair. However, connecting to more storage nodes may lead to larger delay. For instance, in Fig.  \ref{Fig.1WirelessDTN},  the new node may  have to wait longer time to meet more nodes which are moving.  Another approach of improving reliability is  to retransmit the lost packets from the surviving nodes to the new node. However, this will lead to larger delay and also feedback may be necessary. In what follows, we propose to increase storing capacity of the nodes to increase the probability of successful regenerating.

\begin{figure}
\centering
\psfrag{n11}[][][1.5]{ $a_1$ }
\psfrag{n12}[][][1.5]{$b_1$ }
\psfrag{n21}[][][1.5]{ $a_2$ }
\psfrag{n22}[][][1.5]{$b_2$ }
\psfrag{n31}[][][1.5]{ $a_1+b_1+a_2+b_2$ }
\psfrag{n32}[][][1.5]{$a_1+2b_1+a_2+2b_2$ }
\psfrag{n41}[][][1.5]{ $a_1+2b_1+3a_2+b_2$ }
\psfrag{n42}[][][1.5]{$3a_1+2b_1+2a_2+3b_2$ }
\psfrag{n51}[][][1.5]{ $5a_1+7b_1+8a_2+7b_2$ }
\psfrag{n52}[][][1.5]{$6a_1+9b_1+6a_2+6b_2$ }
\psfrag{m1}[][][1.5]{ node 1 }
\psfrag{m2}[][][1.5]{ node 2 }
\psfrag{m3}[][][1.5]{ node 3 }
\psfrag{m4}[][][1.5]{ node 4 }
\psfrag{m5}[][][1.5]{ node 5 }
  \psfrag{f1}[][][2.5]{ $f_1=a_1+2b_1$ }
\psfrag{f2}[][][2.5]{ $f_2=2a_2+b_2$ }
\psfrag{f3}[][][2.5]{ $f_3=4a_1+5b_1+4a_2+5b_2$ }
\psfrag{sec}[][][1.5]{ second transmission }
\psfrag{inf}[][][1.5]{ $\infty$ }
\psfrag{data1data1data1}[][][1.5]{ $d=7,d^{'}=9,\gamma=3.6$ }
\psfrag{data2data2data2}[][][1.5]{ $d=5,d^{'}=6,\gamma=4$ }
  \psfrag{ylabel}[][][1.5]{ $p_s$ }
\psfrag{xlabel}[][][1.5]{ $p_c$ }
\resizebox{8cm}{!}{\epsfbox{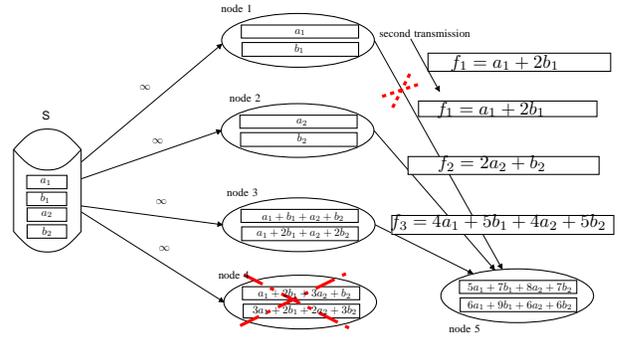}}
\caption{Retransmission is used to recover the lost packet.}
\label{FigExampleAddStorage_a}
\end{figure}

\begin{figure}
\centering
\psfrag{n11}[][][1.5]{ $a_1$ }
\psfrag{n12}[][][1.5]{$b_1$ }
\psfrag{n21}[][][1.5]{ $a_2$ }
\psfrag{n22}[][][1.5]{$b_2$ }
\psfrag{n31}[][][1.5]{ $a_1+b_1+a_2+b_2$ }
\psfrag{n32}[][][1.5]{$a_1+2b_1+a_2+2b_2$ }
\psfrag{n41}[][][1.5]{ $a_1+2b_1+3a_2+b_2$ }
\psfrag{n42}[][][1.5]{$3a_1+2b_1+2a_2+3b_2$ }
\psfrag{n51}[][][1.5]{ $5a_1+7b_1+8a_2+7b_2$ }
\psfrag{n52}[][][1.5]{$6a_1+9b_1+6a_2+6b_2$ }
\psfrag{m1}[][][1.5]{ node 1 }
\psfrag{m2}[][][1.5]{ node 2 }
\psfrag{m3}[][][1.5]{ node 3 }
\psfrag{m4}[][][1.5]{ node 4 }
\psfrag{m5}[][][1.5]{ node 5 }
\psfrag{f1}[][][2.5]{ $f_1=a_1+2b_1$ }
\psfrag{f2}[][][2.5]{ $f_2=2a_2+b_2$ }
\psfrag{f3}[][][2.5]{ $f_3=4a_1+5b_1+4a_2+5b_2$ }
\psfrag{sec}[][][1.5]{ second transmission }
  \psfrag{flink}[][][2.5]{ $f_1+3f_2+2f_3$ }
   \psfrag{flink2}[][][1.5]{ $f_1+3f_2+2f_3$ }
\psfrag{inf}[][][1.5]{ $\infty$ }
\psfrag{data1data1data1}[][][1.5]{ $d=7,d^{'}=9,\gamma=3.6$ }
\psfrag{data2data2data2}[][][1.5]{ $d=5,d^{'}=6,\gamma=4$ }
  \psfrag{ylabel}[][][1.5]{ $p_s$ }
\psfrag{xlabel}[][][1.5]{ $p_c$ }
\resizebox{8cm}{!}{\epsfbox{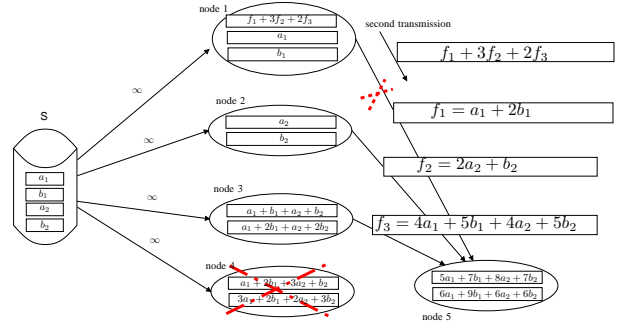}}
\caption{Larger storage capacity increases the probability of successful regeneration than retransmission.}
\label{FigExampleAddStorage_b}
\end{figure}

For illustration, consider an example of a distributed storage system with four nodes shown in Fig. \ref{FigExampleAddStorage_a}. Assume $M=4$, $n=4$, $k=2$, $d=3$ and each node stores two fragments ($\alpha=2$). By (\ref{Eq-MSR}), the optimal repair-bandwidth  in a lossless network is  $\beta=1$. Then, when a node fails, the new node is regenerated by acquiring $\beta=1$ fragment (linearly encoded from data stored in surviving node and denoted by $f_1$, $f_2$ and $f_3$ respectively) from each of three surviving nodes. We assume the channels with the erasure probability $\varepsilon=0.1$. We also assume no feedback to surviving nodes. Thus the retransmission of fragment $f_1$ gives the probability of success $p_s=(\sum_{i=1}^{2}(1-\varepsilon)^i \varepsilon^{(2-i)})(1-\varepsilon)^2=0.8019$. Yet if  a linear combination of  $f_1, f_2$, and $f_3$ can be stored in node $1$, as shown in Fig. \ref{FigExampleAddStorage_b}, then the probability of success is $p_s=\sum_{i=3}^{4}(1-\varepsilon)^i \varepsilon^{(4-i)}=0.9477$ (since any $3$ out of $4$ transmitted fragments can regenerate the new node). Note that two schemes use the same bandwidth for repair. The example shows that increasing storage space can increase the probability of sucessful repair. In the following, we formulate this problem in a general setting.

\begin{figure*}
\centering
\psfrag{node1}[][][1.0]{ node $1$ }
\psfrag{node2}[][][1.0]{ node $2$ }
\psfrag{noden}[][][1.0]{ node $n$ }
\psfrag{a}[][][1.0]{ $\infty$ }
\psfrag{vp}[][][1.5]{ $\vdots$ }
\psfrag{ip}[][][1.5]{ $\ddots$ }
\psfrag{a1}[][][1.0]{ $\alpha_1$ }
  \psfrag{a2}[][][1.0]{ $\alpha_2$ }
   \psfrag{beta1}[][][1.0]{ $\beta_1$ }
    \psfrag{beta2}[][][1.0]{ $\beta_2$ }
    \psfrag{a3}[][][1.0]{ $\alpha_1+\alpha_2$ }
    \psfrag{eq}[][][1.0]{ $\Rrightarrow$ }
    \psfrag{inf}[][][1.0]{ $\infty$ }
     \psfrag{d1}[][][1.0]{ $d_1$ }
      \psfrag{d2}[][][1.0]{ $d_2$ }
       \psfrag{Cut}[][][1.0]{ cut }
\resizebox{14cm}{!}{\epsfbox{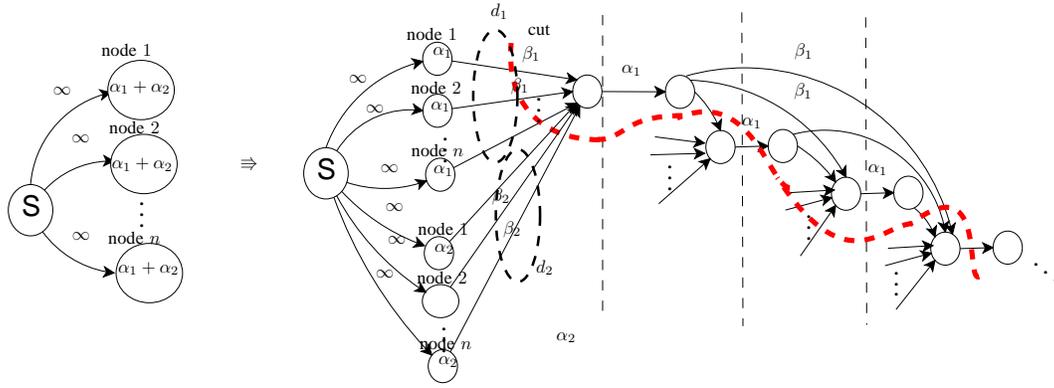}}
\caption{Cut analysis for the network with additional storage space $\alpha_2$ which can increase $p_s$.}
  \label{AddStorageAnalysis}
\end{figure*}

Consider the repair process in a distributed storage system where each node stores $\alpha=\alpha_1+ \alpha_2$ fragments. Assume $k \alpha_1$ fragments from $k$ nodes can rebuild the original file. If channels are error-free then a new node is regenerated by  transmitting $\beta_1$ fragments encoded from $\alpha_1$ stored data of a node. To increase the probability of successful regenerating, each surviving node transmits $\beta_2$ additional fragments to the new node. To simplify cut analysis on the network, we split $\alpha_1$ and $\alpha_2$ storage fragments of each node, as shown in Fig. \ref{AddStorageAnalysis} (it can be proved that there is no loss in separating $\alpha_1$ and $\alpha_2$ ). Note that $\beta_1$ fragments are encoded from $\alpha_1$ stored fragments, and $\beta_2$ fragments are encoded from other $\alpha_2$ stored fragments. In total, $d \beta_1+d \beta_2$ fragments are transmitted from surviving nodes. However, due to link failures, the new node receives $d_1 \beta_1$ and $d_2 \beta_2$   fragments where $d_1, d_2 < d$. We shall show how to calculate $p_s$ from $d_1, d_2, \beta_1, \beta_2$ as follows.

\begin{pro} \label{Proposition:beta2bound}
To regenerate a new node, which receives from $d$ storage nodes by $\beta_1$ fragments from $\alpha_1$ fragments of each surviving node and $\beta_2$ fragments from other $\alpha_2$ fragments of the node,  then the new node can be regenerated by linear network coding if $d_1\beta_1+d_2\beta_2+\sum_{i=1}^{k-1}min(\alpha_1,(d-i)\beta_1) \geq M$.
Here $d_1$ and $d_2$ are the number of links  in the network for transmitting  $\beta_1$ and  $\beta_2$ fragments without erasure, respectively.
\end{pro}


Then,  the probability of successful repair is,
\begin{equation}
 p_s=\sum_{d_1=0}^{d} \sum_{d_2=0}^{d}\binom{d}{d_1} \binom{d}{d_2} (1-\varepsilon)^{d_1+d_2} \varepsilon^{2d-(d_1+d_2)}I(d_1,d_2)
 \end{equation}
where,
\begin{equation} I(d_1,d_2)=
 \begin{cases}
 1   &\text{if  }   d_1\beta_1+d_2\beta_2 \\
 & \geq    M-\sum_{i=1}^{k-1}min(\alpha_1,(d-i)\beta_1), \\
 0 & \text{otherwise  }
 \end{cases}
 \label{I_function}
 \end{equation}

Proof: Own to space limitation, we skip the proof here.
\begin{figure}
\psfrag{xlabel}[][][2.0]{ $\gamma_{th}$ }
\psfrag{ylabel}[][][2.0]{ $\alpha_{th}$ }
\psfrag{title}[][][1.0]{Maximizing probability of successful regenerating. The system has $n=10, k=5, d=9$. }
\resizebox{8cm}{!}{\epsfbox{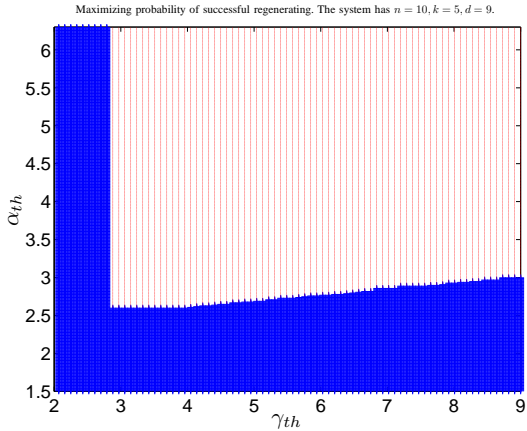}}
 \caption{The filled blue area is for the MSR code providing a higher probability of successful regenerating than the MBR code.}
  \label{MSRorMBR}
\end{figure}

By the Proposition \ref{Proposition:beta2bound}, we can find the optimal storage allocation to achieve the maximum probability of successful regenerating for a given storage space and repair bandwidth. More formally, assume that storage space per node is bounded by $\alpha_{th}$ and the maximum possible total repair-bandwidth is $\gamma_{th}$. Then the optimization storage allocation  is formulated as,

\begin{equation}
 \begin{array}{lc}
 \max\limits_{\alpha_1,\alpha_2} &  p_s \\
 \mbox{\text{s. t.}} & \alpha_1+\alpha_2 \leq \alpha_{th}   \\ & d\beta_1+d\beta_2 \leq \gamma_{th}  \\ & \beta_2 \leq \alpha_2.
 \end{array}
 \label{opt-max}
 \end{equation}

We use an example for illustration, which finds the optimal storage allocation between MSR and MBR codes in a network with $n=10, k=5$ and $\varepsilon=0.1$. We know that MSR codes use less storage, but they need larger repair-bandwidth. On the other hand, MBR codes use more storage at each storage node, but they have lower repair-bandwidth. We use the storage and bandwidth as variables, as shown in Fig. \ref{MSRorMBR}. Then  we observe that for low storage space, the optimal solution is to use MSR codes. It is interesting  to see that for the large storage and low bandwidth region, the optimal solution is still to use MSR codes. Our investigation shows that MBR codes lead to higher probability of successful regeneration only when there is high storage space and large bandwidth available.

\section{Conclusions} \label{sec:conclusion}

We study the regeneration problem for distributed storage systems when the channels are unreliable.
We show the fundamental storage-bandwidth tradeoff for erasure networks. Then the probability of successful regenerating is analyzed.  We use two approaches to maximize the probability of successful regenerating. The first  is to use more surviving nodes for repair. The second is to increase the storage space of storing nodes.  We show that the optimal solution relies on the channel-erasure probabilities. We show that  increasing storage space can also increase reliability for repair.

\end{document}